# Doping of large amount tetravalent Ge ions into $Fe_2O_3$ structure and experimental results on modified structural, optical and electronic properties

Divya Sherin G T[a] and R.N Bhowmik[a*]

[a] Department of Physics, Pondicherry University, Kalapet-605014, India.

[*]Corresponding author: rnbhowmik.phy@pondiuni.edu.in

Abstract

We report the doping high concentration of tetravalent $Ge^{4+}$ ions (5 mol % for $x$ = 0.05 to 30 mol % for $x$ = 0.30) at the $Fe^{3+}$ sites of $Fe_{2-x}Ge_xO_3$ system by chemical coprecipitation route. The charge state of Fe and Ge ions has been modified into lower values, in addition to their normal +3 and +4 states, to stabilize the rhombohedral phase of hematite ($\alpha$-$Fe_2O_3$) structure. X-ray photoelectron spectra and optical band gap measurements indicated a combination of ionic and covalence character of metal-oxygen bonds as an effect of Ge doping in hematite structure. The Ge doped hematite samples have exhibited wide band gap semiconductor properties with band gap 4.50-4.70 eV and remarkably enhanced electrical conductivity ($\sigma \sim 10^{-4}$ S/m) in comparison to $\alpha$-$Fe_2O_3$ ($10^{-11}$ S/m). The thermo-conductivity measurements using warming and cooling modes showed a highly irreversible feature in the semiconductor regime at higher temperatures. Some of the samples indicated metal-like state at lower temperature, in addition to semiconductor state. Our experimental results confirmed the strategy of enhancing electrical conductivity by doping tetravalent ions in hematite structure. It has been understood that combination of ionic and covalence character of the metal-oxygen bonds has played an important role in modifying the semiconductor properties in Ge doped $Fe_2O_3$ system.

1. Introduction

Hematite is a n-type semiconductor with optical band gap of 2.10-2.2 eV [1, 2]. The α-$Fe_2O_3$ stabilizes in rhombohedral structure (R-3c space group) alternatively occupied by Fe and O ions in the hexagonal planes. It has promising applications in the field of Lithium-ion batteries [3,4], supercapacitor, gas sensing [5], and as contrast agents in MRI [6]. The performance of hematite is hindered by its low carrier mobility ($10^{-2}$ $cm^2$ $V^{-1}$ $s^{-1}$), high electron-hole recombination rates (~ 6 ps) [7] and high resistivity ($10^{12}$ Ω.m) [8]. Further, the electron charge transport is highly anisotropic in its Rhombohedral crystal structure (conductivity along out of plane direction is nearly 4 times smaller than the in-plane conductivity). This leads to continuous attempts for enhancing semiconducting properties with modified optical band gap and enhancement of electrical charge transport in α-$Fe_2O_3$ structure by doping divalent (Ni, Zn, Cd) [9-10] and trivalent (Al, Cr, Ga) [11-14] metal ions.

Recently, density functional theory (DFT) calculations and experimental results have suggested that doping of tetravalent (Si, Sn, Ti, Ge) ions in the α-$Fe_2O_3$ structure could be an alternative strategy for achieving high electrical conductivity [15-20]. The doping of tetravalent ions favours the formation of stable rhombohedral structure with n-type semiconductor properties. The charge carrier density in tetravalent metal doped hematite system strongly depends on polaron formation and threshold limit of doping depends on the tetravalent ions. Interestingly, heavier doping of tetravalent ions causes strong modification in electronic structure. It induces new electronic levels between the valence band and Fermi level, as noted for Sn doping ≥16.7 atm % in hematite [16], or causes multipole clustering of polarons after a critical dopant (Ge, Sn, Ti) level [18]. The large amount of doping of $Ge^{4+}$ ions into hematite is important for its applications in spintronic devices due to the expectation of low resistivity (60 Ω.cm), high carrier mobility (3800 $cm^2Vs^{-1}$ for electrons, 1800 $cm^2V.s^{-1}$ for holes), high spin-orbit coupling (290 meV) [21]. Rodrigues et al. [22] demonstrated that Ge doping in hematite enhanced electronic conductivity by increasing carrier mobility and delocalizing the charge carriers by interfering with small polaron formation. The chemical route has been mainly adopted for incorporation of $Ge^{4+}$ ions into $\alpha$-$Fe_2O_3$ structure. However, solubility limit of the Ge ions in hematite has been reported to be low, typically 5-10 % [20, 22]. The grain morphology depends on solubility limit. Hydrothermal method formed circular nanosheets at 2 at % Ge doping and assembled nanosheets at 5 at % Ge doping [23]. The synthesis of $Fe_2O_3$:Ge nanofibers at 10 % of Ge doping using electrospinning route showed impurity maghemite ($\gamma$-$Fe_2O_3$) phase, but the material found to be an excellent candidate for charge storage applications [24].

In this work, we have attempted to dope tetravalent $Ge^{4+}$ ions from 5 mol % to 30 mol % into the lattice structure of $\alpha$-$Fe_2O_3$ system. Primary interest is to confirm the stabilization of rhombohedral phase at low and high dose of doping and provide a detailed information of structural parameters. X-ray photoelectron spectroscopy has been used for studying surface chemical composition and charge state of the Fe and Ge ions. The second interest is to report the modified semiconductor properties, including enhancement of electrical conductivity and optical band gap, as an effect of wide range of Ge doping in hematite structure.

2. Experimental techniques

2.1 Material preparation

The Ge-doped $Fe_2O_3$ system ($Fe_{2-x}Ge_xO_3$) with compositions of $x$ = 0.05, 0.1, 0.20 and 0.30 were prepared by chemical co-precipitation route. The $FeCl_2.4H_2O$ (99.99% pure) and $GeCl_4$ (99.9999% pure) were used as precursors. Initially, 3.090 g of $FeCl_2$ (for $x$ = 0.05) was mixed with distilled water and concentrated HCl was added drop by drop while temperature of the solution was maintained at 60 °C (± 5 °C). The colour of the solution was changed from hazy yellow to clear sunset orange and confirmed the transformation of $FeCl_2$ into $FeCl_3$. Then, measured amount of $GeCl_4$ (0.071 mL for $x$ = 0.05) was mixed into the prepared $FeCl_3$ and the solution temperature was maintained at 85 °C (± 5 °C) during chemical reaction. A precipitating agent NaOH was added to maintain the pH value at approximately 6.60 to obtain the maximum precipitation and less agglomeration among the nanoparticles. The slightly acidic pH value was maintained to prevent significant reduction of $Fe^{3+}$ into $Fe^{2+}$ ions which aid in formation of prominent secondary phases, such as FeO and $Fe_3O_4$ [25]. The temperature was constantly monitored using a thermometer. The entire procedure was performed in air. After an hour of stirring, the precursors of the solution co-precipitated together. A reddish-brown co-precipitate was washed multiple times using deionized water and dried in a box furnace at 200 °C for 2 hours (as-prepared sample). In similar manner, measured amounts of $FeCl_2$ (2.985 g for $x$ = 0.1, 2.798 for $x$ = 0.20, 2.619 g for $x$ = 0.30) and $GeCl_4$ (0.142 mL for $x$ = 0.1, 0.2280 mL for $x$ = 0.20, 0.425 mL for $x$ = 0.30) were used to prepare compositions for Ge concentrations 10 mol %, 20 mol % and 30 mol %. The as-prepared samples were made in the form of pellets of diameter 10 mm and ~1 mm thickness by applying pressure of 4. One set of the pellets was heat treated in air at 550 °C for 6 hours to ensure rhombohedral phase of α-$Fe_2O_3$ [26]. The heat treated samples are henceforth denoted as Ge5, Ge10, Ge20 and Ge30 for Ge doping concentration $x$ = 0.05, 0.1, 0.20 and 0.30, respectively. The pellet shaped samples were used for structural characterization and study of optical and electrical properties.

2.2 Material characterization

X-ray diffraction (XRD) patterns of the samples were recorded at room temperature by employing *Rigaku SmartLab SE* using $CuK_α$ radiation ($λ$ = 1.5406 Å) in the 2θ range 20-70 ° with step size of 0.01 °. Raman spectra were recorded at room temperature using the confocal scanning spectrometer (ReniShaw, UK) configured with 785 nm diode laser source at an optimized power of 5 mW and data acquisition time for 60 seconds in the range of 100-2000 $cm^{-1}$. The elemental composition and charge states of the ions can be obtained from x-ray photoelectron spectra (XPS). The XPS were recorded using Thermo Scientific instrument (K-Alpha-KAN9954133 model) with a monochromatic Al-K$_\alpha$ source. The survey scans were recorded in the binding energy range of 0-1350 eV with a spot size of 400 μm and step size of 1.0 eV. The narrow scans of Fe 2p, Fe 3p, Fe 3s, Ge 3p, Ge 3s, Ge 3d bands were recorded with a step size of 0.1 eV. UV-Vis-NIR spectrometer (Shimadzu, 3600) was used to record optical absorption spectra in the range of 200-800 nm$^{-1}$ and to calculate optical band gap of the Ge doped hematite system. The temperature dependent current-voltage (I-V) characteristics were recorded in the temperature range of 303 K to 623 K by sweeping the voltage from 0 to 50 V in both warming and cooling modes, using a high resistance meter (6517B, Keithley USA). The pellet shaped samples were coated with graphite pencil lead to ensure good electrical contact between sample and electrodes (capacitor mode), where a sample was sandwiched between two Pt electrodes placed and the sample holder was placed in a home-made heater for temperature control.

3. Results and discussion

3.1 X-ray diffraction

The chemical routed as-prepared sample exhibited amorphous phase (not shown in Figure) with mild signatures of (104) and (110) planes corresponding to α-Fe$_2$O$_3$ structure. The heat treatment of the chemical routed samples at 550 °C in air has formed polycrystalline phase (Figure 1(a-d)). The reflections corresponding to (012), (104), (110), (113), (024), (116), (122), (214), (300) planes are consistent to rhombohedral (R-3c space group) structure of α-Fe$_2$O$_3$ [25-27]. The Ge doped hematite structure in the present work has been stabilized at comparatively low temperature [21], without any significant impurity phase [24], in comparison to single phase stabilization of α-Fe$_2$O$_3$ structure by heat treatment of coprecipitated sample (pH = 3) at 800 °C [25]. Rietveld refinement of the XRD patterns has been performed using *FULLPROF* software (version: September 2020) and a refined structure with atomic positions in rhombohedral phase has been shown in Figure 1 (e). The XRD patterns have been best fitted using pseudo-voigt shape. The occupancy of Fe/Ge (0, 0, $z_{Fe}$) and O (0, $y_O$, 0.25) atoms, lattice parameters (a,b,c), effective coordination number ($N_{eff}$) of octahedral Fe ions, bond length ($d_{Fe-O}$) and bond angles ($<Fe-O-Fe$), Fe-O bond distortion ($\Delta$) and degree of distortion ($D_{Fe-O}$) are shown in Table 1. The chemical compositions of the samples from XRD pattern are consistent with the expected value with small variation in oxygen content. The atoms in the refined structure showed displacement from their normal positions (0, 0, $z_{Fe}$ ~ 0.3511) for Fe/Ge and (0, $y_O$ ~ 0.3225, 0.225) for O ions [28]. The *a, b, c* values are found to be within the range of Rhombohedral phase of α-$Fe_2O_3$. Considering a slightly smaller ionic radius of $Ge^{4+}$ (0.54 Å) than that of $Fe^{3+}$ (0.64 Å), the lattice parameters are expected to be smaller in Ge doped samples. However, lattice parameters *a(b), c* and cell volume (V) in the Ge doped samples showed an increasing trend as a function of Ge doping (*x*) and lattice parameters slightly decreased for *x* = 0.30. Subsequently, mass-volume density of the samples decreased with Ge20 exhibited higher density. But the *c/a* ratio has been maintained at 2.73, which is a typical character for Rhomboheral phase, for all the Ge doped samples. At the local level, displacement of ions in the Fe-O bonds ($\Delta$ ~ 35-58 x $10^{-4}$ Å) affects the variation in lattice parameters, $N_{eff}$ for octahedral sites of Fe ions, $d_{Fe-O}$, $<Fe-O-Fe$ and $D_{Fe-O}$ [34]. The $N_{eff}$ value has significantly increased with Ge doping (6.34-8.26) with reference to the normal value 6 for octahedral environment of Fe sites in hematite structure. A considerable modification in the size and shape of $FeO_6$ octahedron due to Ge doping [29] can be realized from the variation of edge sharing (E) and face sharing (F) Fe-O bond lengths ($d_{Fe-O(E)}, d_{Fe-O(F)}$) in Ge5 (2.16 Å, 1.91 Å), Ge10 (2.16 Å, 1.91 Å), Ge20 (2.20 Å, 1.89 Å) and Ge30 (2.15 Å, 1.91 Å) samples in comparison to pristine hematite (2.11 Å, 1.94 Å [28]). Similarly, Fe-Fe distances in the octahedron along edge shared and face shared ($d_{Fe-Fe(E)}$ and $d_{Fe-Fe(F)}$) have shown a significant variation as an effect of Ge doping. This anisotropic nature of the Fe-O/Fe-Fe bond lengths controls the <Fe-O-Fe bond angles and preferential orientation of the crystallites along the *a-b* plane or *c*-axis. The degree of distortion ($D_{Fe-O} = \frac{d_{Fe-O(E)}}{d_{Fe-O(F)}} -$ 1 [34]) in the bond lengths has been found 12-16 % with higher distortion for $x = 0.2$ sample. The lattice distortion and variation of $N_{eff}$ indicated local symmetry reduction by high level doping of Ge ions in hematite structure. This has been understood from the intensity ratio ($I_{(110)}/I_{(104)}$) between (110) (along c-axis, face sharing $d_{Fe-Fe}$) and (104) (edge sharing $d_{Fe-Fe}$) planes [31]. The normalized intensity ratio $I_{(110)}/I_{(104)}$ (~ 0.45, 0.76, 0.85 and 0.56 for Ge5, Ge10, Ge20 and Ge30, respectively) implies that Ge doping favours growth along (110) plane direction till $x = 0.20$ beyond which crystal growth along (110) plane relatively slowed down.

The crystallite size ($D$) and microstrain ($\varepsilon$) have been estimated from the Williamson-Hall (W-H) plot (Figure 1(g-j)) using the equation [30],

$$\beta \cos\theta = \left(\frac{K\lambda}{D}\right) + 4\varepsilon \sin\theta \qquad (1)$$

A linear fit of the plot $\beta\cos\theta$ vs $4\sin\theta$ was used to obtain the crystallite size ($D = k\lambda/intercept$) and micro-strain ($\varepsilon$ = slope) values. The coexistence of positive and negative slopes indicated that metal-oxygen (M-O) bonds have undergone simultaneously expansion and compression to accommodate the tetravalent $Ge^{4+}$ ions in hematite structure [29]. The average crystallite size ($D_{avg}$) and microstrain ($\varepsilon_{rms}$) values are shown in Table 1. The samples have shown a general tendency of decreasing $D_{avg}$ and increase of negative $\varepsilon_{rms}$ with the increase of Ge doping, except notable discrepancy for Ge20 sample. The particle size distribution curve using ImageJ software on FE-SEM image of the Ge20 sample revealed a peak around 11.9 nm (refer supplementary information), which is slightly smaller than average crystallite size (~ 35 nm). This means clustering of small particles in the samples. The polydispersity index ($PdI = \left(\frac{\sigma}{d}\right)^2$, where σ is standard deviation of mean particle diameter (d)) value of 0.045 indicated polydispersive nature in the sample.

3.2 Raman spectra

Complementary microscopic information for local level disorder in Ge doped samples was obtained by analysing Raman spectra. The as-prepared samples showed undefined Raman spectra (Figure 2(a)) due to amorphous structure of the as-prepared samples, although the presence of short-range structural ordering has been indicated by the appearance of some kinks. The broad peaks in as-prepared Ge5asp, Ge10asp, and Ge20asp samples are resolved into distinct peaks for the sample with highest doping ($x = 0.3$). This is an indication of better crystalline phase stabilization at lower heat treatment temperature for the x = 0.3 sample. DFT calculations [33] also provide evidence on the favoured incorporation of Ge into hematite structure than Si and Sn owing to its enthalpy for formation and ionic radius. The heat treatment of the as-prepared samples at 550 °C produced well defined crystalline phase (XRD patterns) and Raman active modes, as predicted by Factor group analysis for rhombohedral crystal structure [32]. The Raman active modes in crystalline samples are marked in Figure 2(b-e) and small variation of the peak positions for different samples can be seen from Table 2. The superposition of $E_g$ (2) and $E_g$ (3) modes at about 297 cm$^{-1}$ arises due to overlap of the metal ion movement and stretching of in-plane M-O bonds in rhombohedral lattice [34]. The $E_g(5)$ mode at ~612 cm$^{-1}$ represents symmetric stretching of in-plane M-O bond and unaltered for all the samples. The infrared (IR) active modes in the range of 661- 663 cm$^{-1}$ ($E_u$ marked as 1LO) and 1319 - 1322 cm$^{-1}$ ($E_{2u}$ marked as 2LO) shows breaking of local symmetry in the lattice structure due to doping-induced disorder [35-37]. The $E_{2u}$ (~1320 cm$^{-1}$) mode, whose frequency almost twice as that of $E_u$ (~ 660 cm$^{-1}$) mode, is identified as two-phonon mode. The normalized peak area of 1LO mode across the compositions varies as Ge30 > Ge5 > Ge10 > Ge20 whereas the peak area of 2LO mode decreases with increasing '$x$' till *0.20* followed by a sudden increase for $x = 0.3$. The relative area ratio of $E_{2u}/E_u$, on the other hand, varies as 6.07, 6.03, 3.09 and 0.44 for Ge5, Ge10, Ge20 and Ge30, respectively. This indicates increasing degree of disorder at local level of lattice structure for high doping of Ge ions and the doping induced disorder is the highest at Ge concentration ($x = 0.3$). Ge20 sample alone shows a peak at 730 cm$^{-1}$, which can be associated to $A_{1g}$ mode due to Ge-O bond stretching in GeO$_6$ octahedra of rutile type structure [38]. A distinct line at ~1584 cm$^{-1}$ for the Ge5 and Ge10 samples confirmed a two-magnon mode (spin wave), which was identified by MJ

Massey et al. [41] at ~ 1525 cm$^{-1}$ in α-Fe$_2$O$_3$. The two-magnon mode is absent for the (Ge20 and Ge30) samples with higher Ge doping. This gives an idea of critical dopant limit below which spin waves can sustain in Ge doped hematite structure [39] and phonon contribution (lattice disorder) dominates at higher Ge doping. The 2-magnon mode observed in Ge5 and Ge10 shows a slight split due to anisotropy caused by dipole-dipole interaction from the half-filled d-orbital nature of Fe$^{3+}$ ions [40] and corresponds to a significant high-frequency shift (blueshift) from the reported 1540 cm$^{-1}$ to 1584-1589 cm$^{-1}$. The position ($\varpi$) of vibrational modes are affected by the changes in Fe-O bond length ($d_{Fe-O}$), force constant (K) and reduced mass ($\mu_{Fe-O}$). They are correlated by the equations: $d_{Fe-O} = \left(\frac{Z_1 Z_2 e^2}{2\pi\varepsilon_0 K}\right)^{1/3}$ and $K = 4\pi^2 c^2 \varpi^2 \mu$, where K is the force constant (N/m), $c$ is the speed of light in vacuum ($c$ = 3 x 10$^8$ m/s), $\varpi$ is the wave number corresponding to peaks obtained in Raman spectra in m$^{-1}$, μ is the reduced mass in kg, Z$_1$ (+3 for Fe and +4 for Ge) and Z$_2$ (-2 for O) depict the charge state of Fe/Ge and O atoms, e is the elementary charge (e = 1.6 x 10$^{-19}$ C), ε$_0$ is the permittivity of free space (ε$_0$ = 8.8 x 10$^{-12}$ m$^{-3}$kg$^{-1}$s$^4$A$^2$). Force constant values are obtained in the range of 38-275 N/m for Raman active modes, ~ 320 N/m for 1LO, ~ 1270 N/m for 2LO and ~ 1835 N/m for 2-magnon modes. The $d_{Fe-O}$ values corresponding to specific E$_g$(5) (symmetric in-plane stretching of Fe-O bond) are comparable to $d_{Fe-O}$ obtained from refinement of XRD patterns. The decrease of FWHM for E$_g$ (5) band (16.5 cm$^{-1}$ for Ge5, 16.0 cm$^{-1}$ for Ge10, 14.4 cm$^{-1}$ for Ge20, 21.9 cm$^{-1}$ for Ge30) indicates enhancement of in-plane lattice order for Ge doping up to $x$ = 0.20. The force constant and bond-length values are nearly same for the samples with higher Ge doping.

3.3 X-ray photoelectron spectroscopy

The survey scan (Figure 3(a-d)) confirmed the presence of Fe, O, Ge ions and signature of adventitious C in all the samples. The surface charge neutrality of the samples was confirmed from C 1s peak at binding energy (BE) ~ 284.80 eV. The peak profiles for selective elemental bands have been deconvoluted using Voigt and Gaussian shapes (shown in Figures 3 and 4). The peak profile parameters (area, position, FWHM) are shown in Table 3. The formation of metal-oxygen (lattice oxygen) bonds (Fe-O/Ge-O) are confirmed by O 1s peak at BE ~ 530-530.5 eV (FWHM ~ 0.50-1.2 eV) [25]. A distinct shoulder in the O 1s spectra (Fig. 3 (e-h)) at ~ 531-532 eV indicated possibility of C-O bond or oxygen vacancies at the surface [41]. The O 1s band showed redshift (shift to lower BE: Ge5 ~ 530.5 eV, Ge10 ~ 530.46 eV, Ge20 ~ 530.11 eV and Ge30 ~ 529.85 eV) Ge doping. M. Lenglet [42] discussed that redshift implies increasing ionic character while the blueshift of O 1s band indicates an increase of covalent character in transition metal oxide systems. Line scan of the Ge 3s and Ge 3d bands (taken at least 7 points across the surface), as shown in Figure 3 (i-k), revealed that the area under the peak is repeatable for each samples, confirming a good degree of surface chemical homogeneity and composition in the samples. The chemical compositions from line scan was found as $Fe_{1.92}Ge_{0.07}O_3$, $Fe_{1.89}Ge_{0.10}O_3$ and $Fe_{1.63}Ge_{0.36}O_3$ for the Ge5, Ge10 and Ge30 samples, respectively. Alternatively, elemental composition ($Fe_{1.78}Ge_{0.22}O_3$) for the Ge20 sample from energy dispersive x-ray (EDX) spectra (refer supplementary information) also matched close to its expected composition $Fe_{1.8}Ge_{0.2}O_3$. The FE-SEM image also indicates a good uniformity in the surface morphology of the sample.

The electronic charge states of the metal ions have been studied by deconvoluting the narrow scan of Fe 2p (Fig. 4 (a-d)), Fe 3s (Fig. 4(e-h)), Ge 3p (Fig. 4 (i-l) and Ge 3s (Fig. m-p)) XPS bands. The peak profile parameters are shown in Table 3. The *p* bands are expected to form doublet (sub-bands) due to a strong spin-orbit (l-s) coupling (SOC, $s = \pm\frac{1}{2}, l = 1$), whereas the *s* bands are expected to form a singlet due to absence of l-s coupling (l = 0). The Fe $2p_{3/2}$ and $2p_{1/2}$ sub-bands are centred around BE ~710.8±0.59 eV (FWHM ~ 0.94-2.2 eV) and ~ 724.31-725.53 eV (FWHM ~ 2.22-5.25 eV) with satellite peaks at ~ 719 eV and 734 eV, respectively. Some of the allowed multiplets (4 and 2 for j = 3/2 and ½, respectively) in $2p_{3/2}$ and $2p_{1/2}$ sub-bands are realized from the multiple peak components in Table 3, whose peak area is proportional to degeneracy factor 2j+1. The Fe energy difference between Fe $2p_{3/2}$ and Fe $2p_{1/2}$ sub-bands (ΔE) is found ~ 13.40-13.78 eV. The energy difference between ($2p_{3/2}$, $2p_{1/2}$) sub-bands and their associated satellites are found in the range of $\Delta E_{3/2}$ ~ 7.56-8.04 eV and $\Delta E_{1/2}$ ~ 7.77 -8.98 eV, respectively in Ge doped samples. The exhibition of Fe $2p_{3/2}$ and $2p_{1/2}$ sub-bands along with distinct satellite peaks and energy difference within the range of reported values $\Delta E$ ~13.12-13.94 eV [31] and $\Delta E_{3/2}$ ~ 8 eV [43] suggest the presence of $Fe^{3+}$ charge states [25, 31, 44]. Table 3 shows that position of the peak components corresponds to $Fe^{3+}$ and $Fe^{+(3-\delta)}$ states in Ge doped α-$Fe_2O_3$ system appears to be shifted in comparison to different charge states of Fe ions in pure α-$Fe_2O_3$ system [45]. An asymmetric shape (shoulder at lower energy side) in both Fe $2p_{3/2}$ and Fe $2p_{1/2}$ sub-bands clearly indicates the coexistence of Fe ions with charge state less than +3. Further, the low energy side peak component of Fe $2p_{3/2}$ sub-band (711.31 eV to 709.64 eV) and Fe $2p_{1/2}$ sub-band (724.62 eV to 722.91 eV) showed a systematic shift to lower energy, implying the increasing fraction of Fe ions with charge state $+(3-\delta)$ with $0 < \delta < 1$ [44]. This is possible by increasing hybridization in metal-oxygen chemical bonds with the increase of Ge doping. Chen et al. [46] used the value of energy difference between main peak and its satellite to quantify the degree of hybridization. In the present samples, both $\Delta E_{3/2}$ and $\Delta E_{1/2}$ ($\Delta E_{3/2}$ ~ 7.56 eV, $\Delta E_{1/2}$ ~ 7.77 eV for Ge5, $\Delta E_{3/2}$ ~ 8.04 eV, $\Delta E_{1/2}$ ~ 8.70 eV for Ge10, $\Delta E_{3/2}$ ~7.73 eV, $\Delta E_{1/2}$ ~ 8.83 eV for Ge20 and $\Delta E_{3/2}$ ~ 7.97 eV, $\Delta E_{1/2}$ ~ 8.98 eV for Ge30) showed an increasing trend with Ge doping, implying an increasing hybridization between valence O 2p and conduction Fe 3d bands. The average (%) for two charge states of Fe ions using peak area components in Fe $2p_{3/2}$ and Fe $2p_{1/2}$ sub-bands has been estimated for Ge5 ($Fe^{+(3-\delta)}$ ~ 27 %, $Fe^{3+}$ ~ 73 %), Ge10 ($Fe^{+(3-\delta)}$ ~ 45 %, $Fe^{3+}$ ~ 55 %,), Ge20 ($Fe^{+(3-\delta)}$ ~ 49 %, $Fe^{3+}$ ~ 51 %) and Ge30 ($Fe^{+(3-\delta)}$ ~ 41 %, $Fe^{3+}$ ~ 59 %) samples. The Fe 3s band also fitted with two components, which is possible due to interactions between core (3s) and unpaired valence (3d) electrons [47] and different chemical and magnetic environment. The peak component at higher and lower energies can be assigned for high (+3) and low (3-δ) charge states. The energy difference ($\Delta E$ ~ 0.63-1.29 eV) between two components of Fe 3s band is significantly smaller than the reported value (~ 5.7 eV) for $Fe^{2+}$ ions [47, 48]. The results are indicative of increasing covalent nature between Fe-O bonds by doping of Ge ions at the Fe sites of α-Fe$_2$O$_3$ structure. Subsequently, the charge state of Ge ions is expected to be different from normal + 4 state.

The charge state of Ge ions has been understood by analysing the Ge 3p, Ge 3d and Ge 3s bands. The doublets 3p$_{3/2}$ and 3p$_{1/2}$ of Ge 3p band have been centered at BE ~ 120-123 eV and BE ~ 124-129 eV with multiplets in Ge 3p$_{3/2}$ sub-band and energy gap between 3p$_{3/2}$ and 3p$_{1/2}$ sub-bands ($\Delta_{Ge3p}$) ~ 4-4.8 eV. These energy parameters suggest that Ge ions may be mostly in Ge$^{4+}$ state [49], although a fraction of Ge ions with lower charge state is indicated from the shift of low energy peak component of Ge 3p$_{3/2}$ sub-band from 123.46 eV to 122.35 eV with the increase Ge doping and split of Ge 3s band [49-52]. The analysis of Ge 3d band (not shown in Figure) showed 3 components (shown in Table 3). The Ge 3d$_{5/2}$ peak component with BE at 32.6-33.3 eV is assigned for + 4 charge state [55] and the peak components at BE ~ 30.50-32.50 eV could be assigned to Ge ions with charge state + (4-δ; 0 <δ < 3) [55-58]. The BE of Ge 3d peak components for +1, +2, +3 and +4 charge states can be assigned at ~ 30 eV, 30.7, 32.4 eV and 32.9 eV, respectively [50, 58]. The % of different charge states of Ge ions has been estimated from the relative area (%) for low energy (31.63-32.22 eV) and high energy (33.13-33.30 eV) components of Ge 3d band and it has been obtained as Ge5 (Ge$^{+(4-\delta)}$ ~ 68 %, Ge$^{4+}$ ~ 32 %), Ge10 (Ge$^{+(4-\delta)}$ ~ 80 %, Ge$^{4+}$ ~ 20 %,), Ge20 (Ge$^{+(4-\delta)}$ ~ 81 %, Ge$^{4+}$ ~ 19 %) and Ge30 (Ge$^{+(4-\delta)}$ ~ 100 %, Ge$^{4+}$ ~ 0 %) samples. The results suggest that the fraction of lower charge states (close to Ge$^{+2}$/Ge$^{+3}$) increases with the increase of Ge doping concentration and the Ge$^{+4}$ appears to be nearly zero for Ge30 sample. The lowering of charge states increases the radius for both Ge (Ge$^{4+}$: 0.54 Å, Ge$^{3+}$: 0.54 Å, Ge$^{2+}$: 0.87 Å) and Fe (Fe$^{3+}$: 64 Å, Fe$^{2+}$: 0.75 Å) may be one of the reasons for increasing trend of lattice parameter (*c*) and cell volume (*V*) with the increase of Ge doping in hematite structure. The split in Ge 3s band is not normally expected due to absence of l-s coupling (for Ge ions in +4 state, the valence band is fully paired (3d$^{10}$) and no unpaired electron). However, split in Ge 3s band suggests interactions between core (Ge 3s) and unpaired valence electron at 4s band [47, 52] in the presence of non-uniform magnetic environment due to different charge states of Fe ions. In addition, the varied values of $N_{eff}$ (6.0-8.5) hint for multiple charge states of Fe and Ge ions [54]. The effect of increasing co-valence character of metal-oxygen (Fe-O/Ge-O) bonds in the Ge doped α-Fe$_2$O$_3$ system has been understood from the study of semiconductor properties (optical band gap and electrical conductivity) of the samples.

3.4 Optical band gap

The absorption spectra (Figure 5(a)) show a dip at ~ 240-245 nm for all samples. All four samples showed a sharp increase immediately after this dip, followed by a relatively slow decay at higher wavelengths. The dip at ~ 200-245 nm in hematite structure arises due to charge transfer between Fe/Ge and oxygen ions and increase of the optical absorbance below the dip arises due to Fe-Fe (d-d orbital) transitions [53]. The optical band gap has been determined by using a linear fit of modified Tauc plot: $(\alpha h\upsilon)^{1/n} = A (h\vartheta - E_g)$, where α is the absorption coefficient and $A$ corresponds to transition probability from the valence band to conduction band. The Tauc plot as shown in Fig. 5 (b-i) has been best fitted by assigning $n$ = 0.5, indicating direct electronic transition from valence to conduction band and $n$ = 2 corresponding to indirect electronic transition. The direct ($E_g$ = 4.5-4.69 eV and 2.3-2.9 eV) and indirect ($E_g$ =1.49-1.70 eV) band gap values for Ge doped hematite samples have been shown in Fig. 5 (b-e) and Fig. 5 (f-i), respectively. The direct band gap values ($E_g$) of Ge doped samples are consistent to incorporation of Ge in α-Fe$_2$O$_3$ structure by chemical reaction of their oxide forms (hematite ~ 2.0-2.4 eV and GeO$_2$ ~ 6.6 eV). However, band gap decreased in the samples with the increase of Ge doping and similar results were reported by Liu et al. [23]. The direct band gap values can be attributed to Burstein-Moss effect [55] where increase of Ge (which is a typical semiconductor of band gap ~ 0.67 eV) concentration causes population of the density of states within the conduction band and decreasing the direct optical band gap. Another possible reason may be presence of a strong p-d hybridization effect from $\Delta E_{1/2}$ (as shown from XPS) [51, 54] between the 2p dominated valence band and 3d dominated conduction band in the Ge-doped samples. On the other hand, indirect band gap in the samples are comparable to the lower range of direct band gap, which arises due to defect induced states between valence band and conduction band. The defect induced states form local equilibrium levels in the forbidden gap and electronic transitions are coupled with lattice phonons. Increasing intensity of ILO mode from Raman spectra confirms the presence of defect induced energy levels. A combination of direct and indirect electronic transitions in Ge doped samples is advantageous for preventing rapid recombination of charge carriers while retaining suitable indirect gap values useful for water splitting in the visible light region ($E_g$ slightly lesser than 2.0-2.4 eV). Hematite doped with tetravalent elements such as $Sn^{4+}$, $Ti^{4+}$, $Si^{4+}$, $Ge^{4+}$ [19, 53, 54, 56] showed $E_g$ in the range of 1.7-2.4 eV, i.e., narrowing of optical band gap (red-shifted with reference to hematite). This is comparable to the low energy side band gap in our samples. In contrast, $E_g$ at the higher energy side in Ge-doped hematite systems is higher than pure hematite system and it suggests increase of ionic character in the Ge doped hematite system [19]. However, increase of covalence character (as seen from XPS analysis) in the samples with the increase of Ge doping can be verified from the decrease of band gap at high energy side. The refractive index (*n*) of our samples has been calculated using Mosse relation, $n^4 E_g = 95\ eV$ [57] and obtained values are in range of 2.39-2.51 eV (direct, lower energy side), 2.12-2.14 eV (direct, higher energy side) and 2.73-2.82 eV (indirect band gap). The refractive indices for the direct band gap values of Ge doped samples are smaller than the reported value *n* = 2.54 [57] for un-doped hematite microparticles indicating faster propagation of light in our nanoparticle systems.

3.5 DC electrical conductivity

The current-voltage (I-V) characteristics (Figure 6(a-d)) confirmed the semiconducting nature in Ge doped hematite samples. The differential conductivity (*dI/dV* vs *V*) curves in the inset of Fig. 6(a, d) confirmed voltage induced increase of charge mobility in Ge doped samples. The dc electrical conductivity (σ) has been calculated from the I-V curves by using the relation $\sigma = \frac{It}{VA}$, where *t* and *A* are the thickness and area of the pellet-shaped samples, respectively. The thermo-conductivity (σ(*T*)) curves at fixed applied voltage has been calculated during warming and cooling mode of measurements in the temperature (T) range 303-623 K. Figure 6 (e-h) showed thermal hysteresis behaviour (irreversibility between warming and cooling paths) in the $\sigma(T)$ curves at 10 V, especially at the higher temperatures. The semiconducting nature with positive temperature coefficient of conductivity dominated at higher temperatures beyond a critical temperature, such as 440 K, 390 K, 378 K, 360 K in case of Ge5, Ge10, Ge20, Ge30 samples, respectively, below which the $\sigma(T)$ curves are either less temperature dependent due to increasing localization of charge carriers around the ionic sites [21, 56] or indicated metal-like signature with negative temperature coefficient of conductivity. The electrical conductivity of the samples at selected temperatures (350 K, 450 K, 550 K) measured at 10 V of the semiconductor regime has been compared in Table 4. The overall conductivity is remarkably enhanced in Ge doped samples in comparison to un-doped α-$Fe_2O_3$ system (typically $10^{-10}$ S/m) [18]. As shown in Fig. 6(i), the electrical conductivity has gradually enhanced up-to x = 0.20 (of the order of $10^{-6}$-$10^{-4}$ S/m) followed by a slight drop in σ (T) for x = 0.30 (order of $10^{-5}$-$10^{-6}$ S/m). At higher concentration of Ge doping, different kind of effect such as clustering of small polarons may increase the localization of the charge carriers in hematite structure [17, 19]. The increasing concentration of unstable $Ge^{3+}/Ge^{2+}$ charge states (acts as electron acceptor [58]) at the highest doping (x = 0.3) can inhibit effective charge transport. Zhao et al. [56] demonstrated that excessive Ge concentration hinders the growth of crystal along (110) plane in α-$Fe_2O_3$ structure and we realized it from the variation of $I_{(110)}/I_{(104)}$ intensity ratio in XRD pattern. It decreases the in-plane electrical conductivity in hematite structure [32].

The presence of metal-like (M) and semiconductor (S) conductivity states have been further understood from the log ($\sigma$ (T)) vs. 1000/T curves (Fig. 7 (a-h)), following Arrhenius law,

$$\sigma\ (T) = Ae^{-(E_A K_B)/T} \qquad (2)$$

where $A$ is a constant, $E_A$ is the activation energy required for transport of charge carriers between lattice sites, $K_B$ being the Boltzmann constant (1.38 x $10^{-23}$ $m^2.kg.s^{-2}K^{-1}$), $T$ the measurement temperature. Two linear regions of positive (metal-like behaviour) and negative (semiconductor regime) slopes have been observed for most of the samples in warming and cooling modes. The activation energy ($E_A$) in the semiconductor and metal-like states have been calculated by using the relation, $E_A = -\frac{(m*2303*K_B)}{(1.6*10^{-19}e)}$, where $m$ is the slope obtained from the Arrhenius plot and shown in Table 4. The $E_A$ in semiconductor state lies in the range of 1.0-2.65 eV [21]. The positive $E_A$ in the semiconductor regime is related to energy required for hopping of charge carriers between cation sites via oxygen ion (superexchange paths Fe-O-Fe, Fe-O-Ge, Ge-O-Ge etc) [32, 59]. The hopping of charge carriers between two Fe/Ge sites with different charge states ($Fe^{+3}/Fe^{3-\delta}$, $Ge^{4+}/Ge^{4-\delta}$) and strength of coupling in distorted Fe-O/Ge-O bonds reorganize the localization of charge carriers [2, 14, 32]. On the other hand, the metal-like state is not due to free charge carriers in the samples. Rather, negative activation energy in the intermediate metal-like state is associated with readjustment of activation energy between low temperature and high temperature semiconductor regimes. The total reorganization energy ($\lambda$) of the system, thus calculated is slightly higher than that (1.03 eV [60]) predicted by the ab initio model for pure α-$Fe_2O_3$. Higher $\lambda$ values are attributed to the energy required for the doped hematite system to reorganize itself to facilitate charge transfer and hence enhance electrical conductivity ($\sigma$).

The charge transport mechanism as a function of applied voltage in the samples, irrespective of the low and high temperature regimes, have been understood by analysing the current density (J(V)) curves at selected temperatures (350 K, 450 K, 550 K) using Schottky and Poole-Frenkel equations [62]. However, J(V) curves of the samples over a wide voltage range have been best fitted with Poole-Frenkel equation: $J = qn_c\mu \left(\frac{V}{t}\right) e^{\frac{-q\varphi_B}{K_BT}+\frac{q\sqrt{\frac{qV}{\pi\epsilon t}}}{rK_BT}}$ (4)

J, $n_c$, μ, r, $\varphi_B$, $K_B$, q, $\varepsilon_r$ denote the current density, dynamic charge density, mobility of charge carriers, constant (taking values 1-2), barrier height, Boltzmann constant, electron charge and the dynamic dielectric constant, respectively. The Poole-Frenkel conduction mechanism, where the charge transport is controlled by bulk material, has been realized by obtaining linear fit of ln (J/V) vs (V)$^{1/2}$ curves. The ln (J/V) vs (V)$^{1/2}$ plots (Fig. 7 (i-l)) have been fitted with different slopes at lower and higher voltage regions for all the samples. The values of $\varphi_B$ lie in the range of 0.16-0.32 eV, 0.03-0.21 eV and 0.01-0.43 eV for 350 K, 450 K and 550 K, respectively. The $\varphi_B$ values for low and high voltage regions do not show much difference. The average barrier height values ($\varphi_B$) as low as 0.01 eV at 550 K has been observed for Ge20 sample, explaining the reason for it possessing highest σ among the other samples. The facilitation of thermal energy ($K_BT$) mediated charge transfer between lattice sites (via Fe-O-Fe) is evidenced from the reduced barrier height with increase in temperature. Lowered $\varphi_B$ facilitates increased conductivity which is an attractive aspect in designing highly efficient photoanodes. It is to be noted that Ge30 possesses negative $\varphi_B$ at 550 K which may indicate easy charge transport between Fe sites but its overall $E_A$ value (1.09 eV) explains the slight reduction of σ value than that for Ge20 sample.

4. Conclusion

The $Fe_{2-x}Ge_xO_3$ system has been successfully stabilized in rhombohedral phase for a wide range of Ge doping (*x*) from 5 mol % to 30 mol %. The chemical compositions of the samples have been verified close to the expected values using XRD patterns, SEM and XPS analysis. The local lattice structure, surface chemical state and physical properties of the Ge doped samples are well affected by the doping level. The local level structural distortion has been confirmed from the increased effective coordination number, degree of distortion, ionic displacements and additional phonon modes in Raman spectra. The blue shifted nature of the 2-magnon mode in samples of low Ge concentration (*x* = 5mol % and 10 mol %) and its absence in highly doped samples (Ge20, Ge30) helps to establish a critical dopant level of Ge in hematite for propagation of spin waves in Ge doped hematite system. The ionic character of $Fe^{3+}$ and $Ge^{4+}$ ions has been modified and covalence character of metal-oxygen bonds has been enhanced due to multivalent charge states of Fe (+3, +3-δ) and Ge (+4, 4-δ) ions. The increase of lattice parameters with increasing Ge concentration can be attributed to increasing fraction Fe and Ge ions with charge states less (and higher ionic radii) than their normal value of +3 and +4, respectively. The current-voltage characteristics and their first derivatives confirmed semiconductor behaviour and voltage induced charge transport are controlled by Poole-Frenkel (trap assisted conduction) mechanisms over a wide range of applied voltages. The lowering of barrier height with temperature, low activation energy for charge hopping, high electrical conductivity in semiconductor regime and coexistence of indirect and direct optical band gap are some of the notable features in Ge doped samples. The reorganization of hopping charge carriers between Fe/Ge sites with different charge states ($Fe^{+3}/Fe^{+(3-\delta)}$, $Ge^{+4}/Ge^{+(4-\delta)}$) and localization of charge carriers at lower temperatures around the distorted Fe-O/Ge-O bonds are responsible for a metal-like state and smaller electrical conductivity. At higher doping concentration ($x = 0.3$), the formation of small polarons reduced electrical conductivity in the system. However, localization of the multi-valence ions within band gap forms defect induced levels and enhanced overall electrical conductivity and reduced optical band gap in Ge doped samples. Such materials can find potential applications as photoanodes and UV sensors.

**Acknowledgement:** The authors acknowledge experimental support from Department of Physics and Central Instrumentation Facility (CIF-PU), Pondicherry University for XRD, XPS, Raman and optical measurements. The authors thankfully acknowledge Dr. S.A. Khan of Material Science Division, Inter-University Accelerator Centre (IUAC), New-Delhi for FESEM and EDX measurements.

Table 1. Parameters obtained from Rietveld refinement of the powder XRD patterns of Ge5, Ge10, Ge20 and Ge30 samples.

| Samples → Parameters ↓ | Ge5 | Ge10 | Ge20 | Ge30 |
|---|---|---|---|---|
| **Atom positions** | | | | |
| Fe: $(0,0,z_{Fe})$ | (0,0,0.3555) | (0,0,0.3544) | (0,0,0.3547) | (0,0,0.3550) |
| O: $(0,y_O,0.25)$ | (0,0.3121, 0.25) | (0,0.3212, 0.25) | (0,0.3302, 0.25) | (0,0.3180, 0.25) |
| Composition | $Fe_{1.93}Ge_{0.06}O_{3.008}$ | $Fe_{1.904}Ge_{0.096}O_{2.99}$ | $Fe_{1.78}Ge_{0.22}O_{2.87}$ | $Fe_{1.75}Ge_{0.25}O_{3.229}$ |
| $\Delta$ (10$^{-04}$ Å) | 38.32 | 37.88 | 57.53 | 34.70 |
| a=b (Å) | 5.0335 (125) | 5.0365 (86) | 5.0379 (80) | 5.0362 (82) |
| c (Å) | 13.747 (496) | 13.759 (337) | 13.754 (367) | 13.754 (335) |
| V (Å$^3$) | 301.653 (149) | 302.26 (104) | 302.33 (65) | 302.11 (96) |
| c/a ratio | 2.73 | 2.73 | 2.73 | 2.73 |
| $\rho$ (g/cm$^3$) | 5.362 | 5.328 | 5.676 | 5.508 |
| $N_{eff}$ | 6.41 | 6.34 | 8.26 | 7.31 |
| $d_{Fe-O}$(E,F) (Å) | 2.163 (12), 1.911 (18) | 2.164 (12), 1.913 (14) | 2.20 (5), 1.89 (9) | 2.156 (6), 1.917 (14) |
| $d_{Fe-Fe}$ (E, F)(Å) | 3.701 (5), 2.876 (8) | 3.709 (4), 2.876 (6) | 3.70 (4), 2.968 (12) | 3.703 (4), 2.888 (6) |
| <Fe-O-Fe | 85.4(5) | 83.3 (5) | 81.9 (4) | 85.8 (5) |
| | 131.1(5) | 130.5(4) | 129.6 (3) | 131.5 (5) |
| | 93.7(4) | 93.2 (4) | 92.7 (3) | 93.8 (4) |
| <O-Fe-O | 79.0 (7) | 80.7 (7) | 81.7 (5) | 78.7 (7) |
| | 163.0 (11) | 164.8 (11) | 165.8 (7) | 162.8 (11) |
| | 102.4 (10) | 101.5 (3) | 87.9 (3) | 90.3 (4) |
| $D_{Fe-O}$ | 0.13 | 0.13 | 0.16 | 0.12 |
| $D_{avg}$ (W-H plot) (nm) | 30.1 | 20.5 | 35.3 | 19.9 |
| $\varepsilon_{rms}$ | 0.035 % | -0.076% | -0.27 % | -0.12% |

Table 2. Raman modes observed for Ge-doped samples and their corresponding Fe-O bond lengths.

| $\bar{\omega}$ cm$^{-1}$ Ge5 | $\bar{\omega}$ cm$^{-1}$ Ge10 | $\bar{\omega}$ cm$^{-1}$ Ge20 | $\bar{\omega}$ cm$^{-1}$ Ge30 | $d_{Fe-O}$ Å Ge5 | $d_{Fe-O}$ Å Ge10 | $d_{Fe-O}$ Å Ge20 | $d_{Fe-O}$ Å Ge30 | K N/m Ge5 | K N/m Ge10 | K N/m Ge20 | K N/m Ge30 | $d_{Ge-O}$ Å Ge5 | $d_{Ge-O}$ Å Ge10 | $d_{Ge-O}$ Å Ge20 | $d_{Ge-O}$ Å Ge30 | Modes [33] |
|---|---|---|---|---|---|---|---|---|---|---|---|---|---|---|---|---|
| 225 | 227 | 229 | 224 | 5.22 | 4.27 | 4.24 | 4.31 | 37.0 | 37.7 | 38.3 | 36.7 | 4.64 | 4.62 | 4.59 | 4.66 | A$_{1g}$ (1) |
| 246 | 245 | 246 | 245 | 4.92 | 4.06 | 4.05 | 4.06 | 44.3 | 43.9 | 44.3 | 43.9 | 4.38 | 4.38 | 4.38 | 4.39 | E$_g$ (1) |
| 294 | 294 | 296 | 294 | 4.37 | 3.59 | 3.58 | 3.59 | 63.3 | 63.3 | 64.4 | 63.3 | 3.89 | 3.89 | 3.87 | 3.89 | E$_g$ (2) + E$_g$ (3) |
| 412 | 412 | 414 | 412 | 3.50 | 2.87 | 2.86 | 2.87 | 124.3 | 124.3 | 124.1 | 124.3 | 3.10 | 3.10 | 3.09 | 3.10 | E$_g$ (4) |
| 499 | 498 | 500 | 498 | 3.08 | 2.53 | 2.52 | 2.53 | 182.4 | 181.7 | 184.2 | 181.7 | 2.73 | 2.73 | 2.73 | 2.73 | A$_{1g}$ (2) |
| 612 | 612 | 612 | 612 | 2.69 | 2.20 | 2.20 | 2.20 | 274.7 | 274.7 | 274.7 | 274.7 | 2.38 | 2.38 | 2.38 | 2.38 | E$_g$ (5) |
| 659 | 661 | 661 | 661 | 2.56 | 2.09 | 2.09 | 2.09 | 318.2 | 321.4 | 321.4 | 321.4 | 2.27 | 2.26 | 2.26 | 2.26 | E$_u$ |
| 1312 | 1316 | 1318 | 1318 | 1.63 | 1.32 | 1.32 | 1.32 | 1261.3 | 1269.0 | 1277.8 | 1277.8 | 1.43 | 1.43 | 1.43 | 1.43 | E$_{2u}$ |
| 1589 | 1584 | - | - | 1.44 | 1.17 | - | - | 1850.2 | 1838.5 | - | - | 1.26 | 1.26 | - | - | 2-magnon |

Table 3. Peak position, area and FWHM (ß) obtained from XPS spectra for the Ge-doped samples.

| Sample→ Core band | Peak Nos. | Ge5 Area (arb. Units) | Ge5 Position (eV) | ß (eV) | Ge10 Area (arb. Units) | Ge10 Position (eV) | ß (eV) | Ge20 Area (arb. Units) | Ge20 Position (eV) | ß (eV) | Ge30 Area (arb. Units) | Ge30 Position (eV) | ß (eV) |
|---|---|---|---|---|---|---|---|---|---|---|---|---|---|
| C 1s | 1 | 136663 | 284.79 | 0.69 | 49367 | 284.59 | 0.89 | 6426 | 284.03 | 0.96 | 58038 | 284.75 | 1.19 |
|  | 2 | 241574 | 285.48 | 2.08 | 41035 | 284.69 | 0.55 | 29011 | 284.43 | 0.61 | 8687 | 288.70 | 1.56 |
|  | 3 | ---- | ----- | ---- | 60573 | 285.22 | 0.89 | 18167 | 284.73 | 0.69 | ---- | ---- | ---- |
| O 1s | 1 | 93098 | 530.50 | 1.14 | 201122 | 530.46 | 1.20 | 248193 | 530.11 | 1.24 | 331345 | 529.85 | 1.09 |
|  | 2 | 84726 | 532.35 | 2.34 | 293115 | 531.99 | 2.70 | 234700 | 531.76 | 2.14 | 213142 | 531.27 | 2.68 |
| Fe 2p | 1 | 10362 | 710.31 | 0.92 | 23431 | 710.28 | 1.00 | 23990 | 709.88 | 0.93 | 51389 | 709.64 | 0.97 |
|  | 2 | 39563 | 711.31 | 2.02 | 112777 | 711.23 | 2.14 | 91344 | 710.93 | 2.12 | 159157 | 710.69 | 1.97 |
|  | 3 | 147049 | 712.88 | 4.61 | 157873 | 712.88 | 3.67 | 207813 | 712.36 | 4.27 | 217115 | 712.34 | 3.41 |
|  | 4 | 118035 | 718.97 | 7.32 | 6859 | 719.48 | 1.83 | 77829 | 719.01 | 5.43 | 19987 | 719.04 | 2.43 |
|  | 5 | 34465 | 724.62 | 2.89 | 45278 | 724.40 | 2.17 | 6566 | 723.15 | 1.10 | 6518 | 722.91 | 0.80 |
|  | 6 | 80914 | 726.98 | 5.29 | 55794 | 726.29 | 3.11 | 61538 | 724.31 | 2.51 | 55020 | 723.91 | 2.18 |
|  | 7 | ----- | ------ | ----- | ----- | ----- | ----- | 50985 | 726.32 | 3.14 | 125947 | 725.53 | 4.00 |
|  | 8 | 27102 | 733.13 | 5.03 | 4809 | 733.96 | 1.88 | 15666 | 734.56 | 4.89 | 17809 | 733.12 | 3.27 |
| Fe 3p | 1 | 14723 | 55.64 | 1.56 | 22136 | 55.62 | 1.39 | 10291 | 54.92 | 1.53 | 32007 | 55.12 | 2.99 |
|  | 2 | 10529 | 56.62 | 1.69 | 36723 | 56.74 | 1.94 | 34130 | 55.87 | 2.00 | 63127 | 56.20 | 2.27 |
|  | 3 | 11988 | 57.99 | 2.44 | 21651 | 58.03 | 3.38 | 22429 | 57.40 | 2.62 | 42844 | 58.27 | 1.59 |
| Fe 3s | 1 | 8875 | 93.49 | 2.52 | 6535 | 93.83 | 1.92 | 2028 | 93.19 | 1.00 | 11296 | 93.26 | 1.82 |
|  | 2 | 4980 | 94.72 | 2.30 | 5943 | 94.46 | 2.16 | 1264 | 94.01 | 1.09 | 12648 | 94.25 | 2.93 |
| Ge 3p | 1 | 6258 | 123.46 | 2.36 | 10959 | 123.45 | 1.52 | 27205 | 122.35 | 7.72 | 11792 | 123.55 | 1.96 |
|  | 2 | 9187 | 124.66 | 1.72 | 7378 | 124.26 | 1.03 | 29918 | 123.80 | 2.15 | 1161 | 123.90 | 0.68 |
|  | 3 | 2134 | 125.88 | 1.55 | 10972 | 125.11 | 1.33 | 51080 | 125.09 | 2.29 | 4549 | 124.62 | 1.45 |
|  | 4 | 9301 | 128.68 | 3.58 | 6606 | 129.19 | 1.77 | 51581 | 128.71 | 3.22 | 3577 | 128.21 | 1.62 |
| Ge 3d | 1 | 2609 | 32.22 | 1.22 | 14552 | 32.22 | 1.23 | 24685 | 31.88 | 1.49 | 9965 | 31.63 | 1.18 |
|  | 2 | 3800 | 32.61 | 1.25 | 14144 | 32.80 | 1.17 | 23895 | 32.50 | 1.25 | 12588 | 32.21 | 1.26 |
|  | 3 | 3054 | 33.13 | 1.30 | 6969 | 33.30 | 1.25 | 11341 | 33.21 | 1.16 | --- | --- | -- |
| Ge 3s | 1 | 2747 | 182.97 | 3.39 | 13349 | 183.81 | 3.17 | 9825 | 182.37 | 2.27 | 2976 | 182.77 | 2.58 |
|  | 2 | 5688 | 185.04 | 2.82 | 6220 | 185.00 | 1.96 | 29554 | 184.51 | 2.70 | 7846 | 184.20 | 2.37 |
|  | 3 | -- | ---- | --- | ---- | ---- | ---- | 12569 | 186.93 | 3.69 | 727 | 188.43 | 2.80 |

Table 4. Electrical conductivity measured at 10 V and electrical activation energies calculated from Arrhenius equation during warming (W) and cooling (C) cycles and parameters calculated according to P-F mechanism.

| Sample | $\sigma$(S/m) at 10 V | Activation energy ($E_A$) | | | | $\lambda$(eV) | Pool-Frenkel mechanism |
| --- | --- | --- | --- | --- | --- | --- | --- |
| | | Warming (W) | | Cooling (C) | | | $\Phi_B$ (eV) |
| | | Region1 $E_A$ (eV) | Region2 $E_A$ (eV) | Region1 $E_A$ (eV) | Region2 $E_A$ (eV) | | |
| Ge5 | 350 K: 2.30E-8<br>450 K: 6.11E-6<br>550 K: 1.90E-5 | 2.02 | -2.16 | 2.65 | -2.23 | W: -0.14<br>C: 0.42 | 350 K: 0.32<br>450 K: 0.19<br>550 K: 0.21 |
| Ge10 | 350 K: 9.49E-8<br>450 K: 3.14E-6<br>550 K: 1.27E-5 | 1.51 | - | 1.56 | - | W: 1.51<br>C: 1.56 | 350 K: 0.16<br>450 K: 0.21<br>550 K: 0.42 |
| Ge20 | 350 K: 1.98E-6<br>450 K: 3.90E-4<br>550 K: 7.67E-4 | 1.64 | -0.51 | 1.58 | -0.60 | W: 1.13<br>C: 0.98 | 350 K: 0.23<br>450 K: 0.03<br>550 K: 0.04 |
| Ge30 | 350 K: 9.66E-7<br>450 K: 5.7E-5<br>550 K: 2.55E-4 | 2.59 | -1.50 | 1.31 | -- | W: 1.09<br>C: 1.31 | 350 K: 0.20<br>450 K: 0.09<br>550 K: -0.02 |

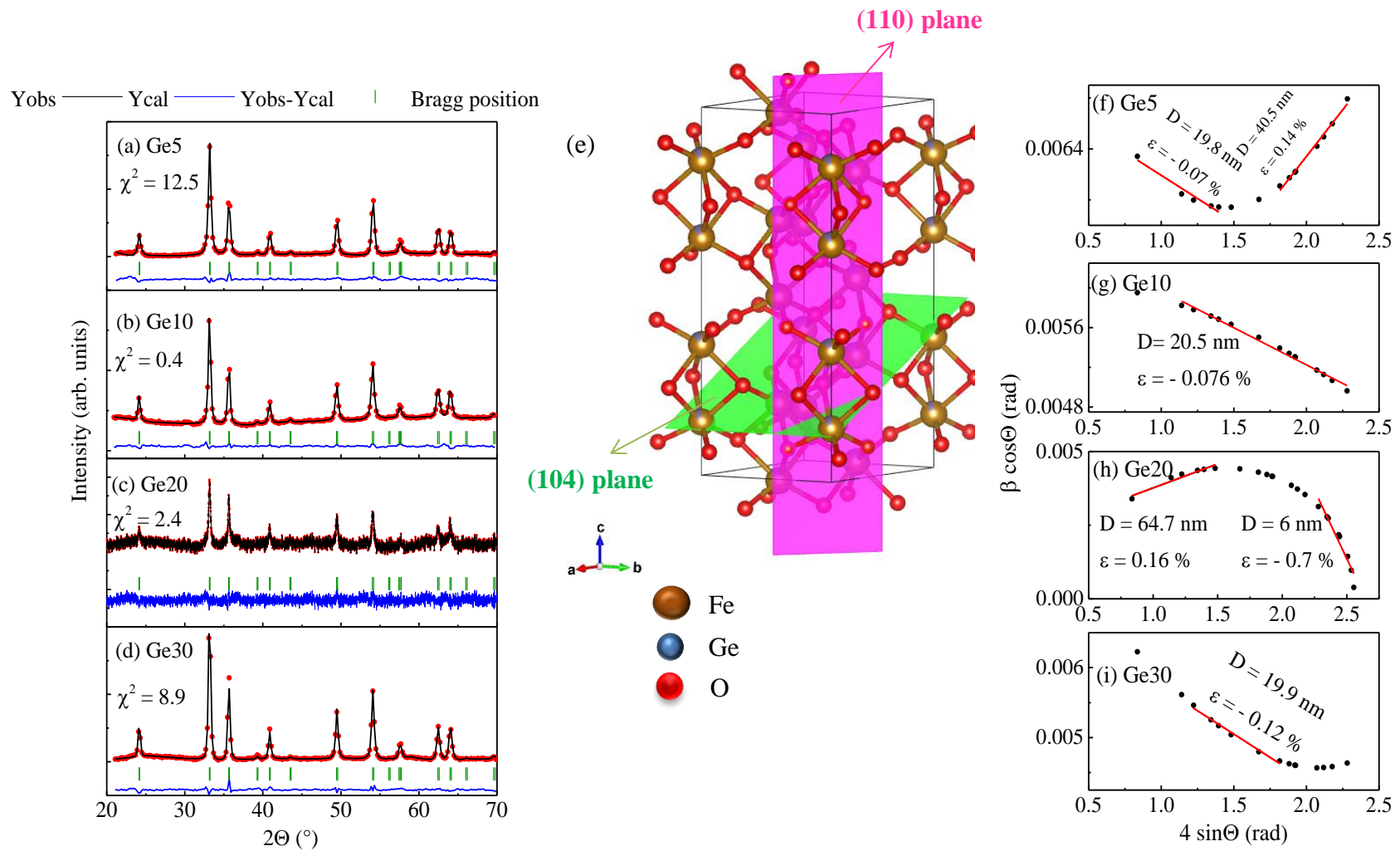

Figure 1. (a-d) Powder XRD patterns of Ge5, Ge10, Ge20 and Ge30 samples after performing Rietveld refinement. (e) Unit cell of Ge30 sample after refinement with (104) and (110) planes shown in green and pink, respectively. (f-i). Linear fits of Williamson-Hall equation (FWHM and 2θ values obtained from refinement) to obtain crystallite size (D) and microstrain (ε).

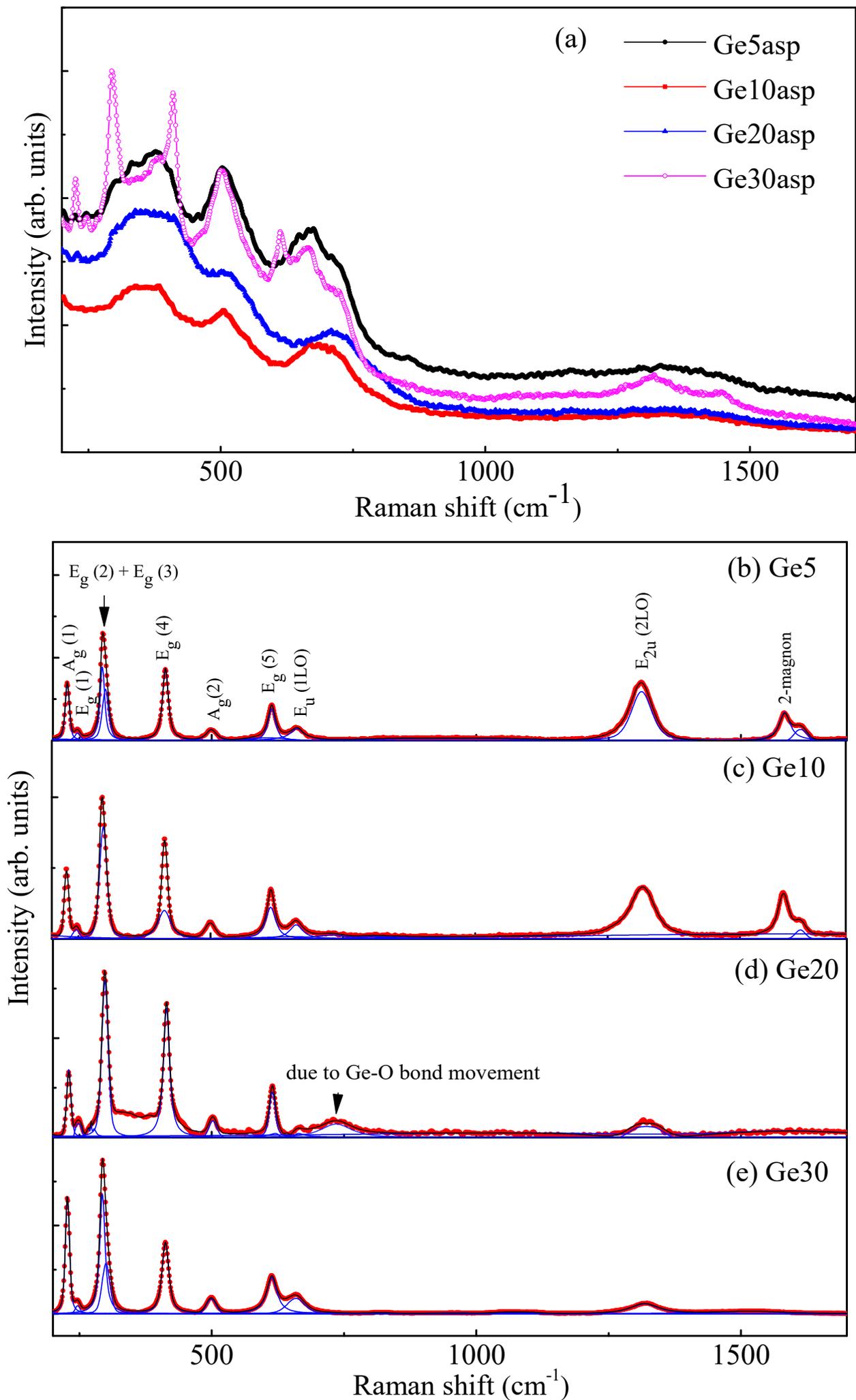

Figure 2. Raman spectra of (a) as-prepared and (b-e) heat treated Ge5, Ge10, Ge20 and Ge30 samples.

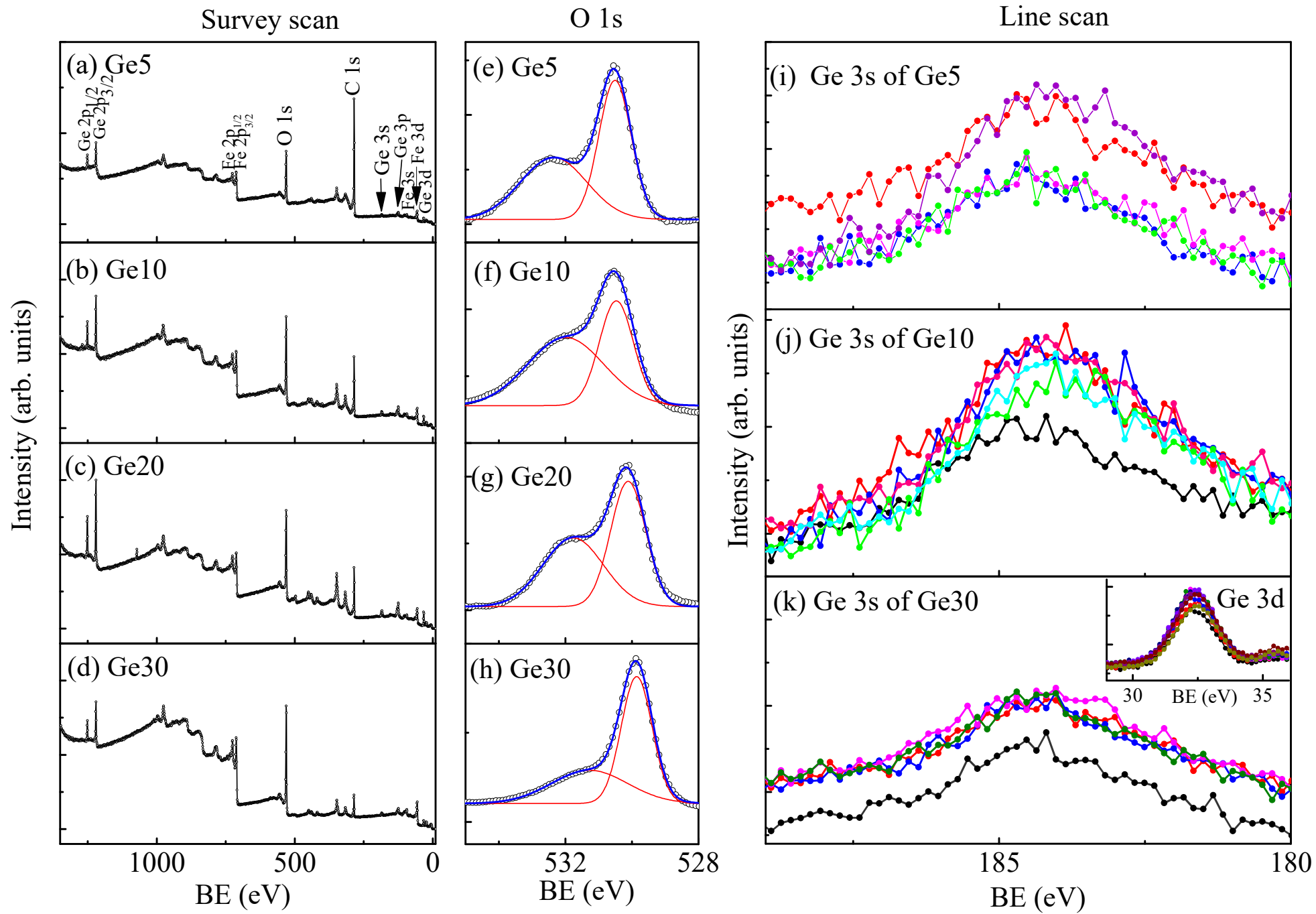

Figure 3. Survey scan (a-d) of Ge5, Ge10, Ge20 and Ge30 samples along with corresponding O 1s (e-h) spectra. Line scan of Ge 3s band (i-k) of the Ge-doped samples (inset showing line scan of Ge 3d band for Ge30 sample).

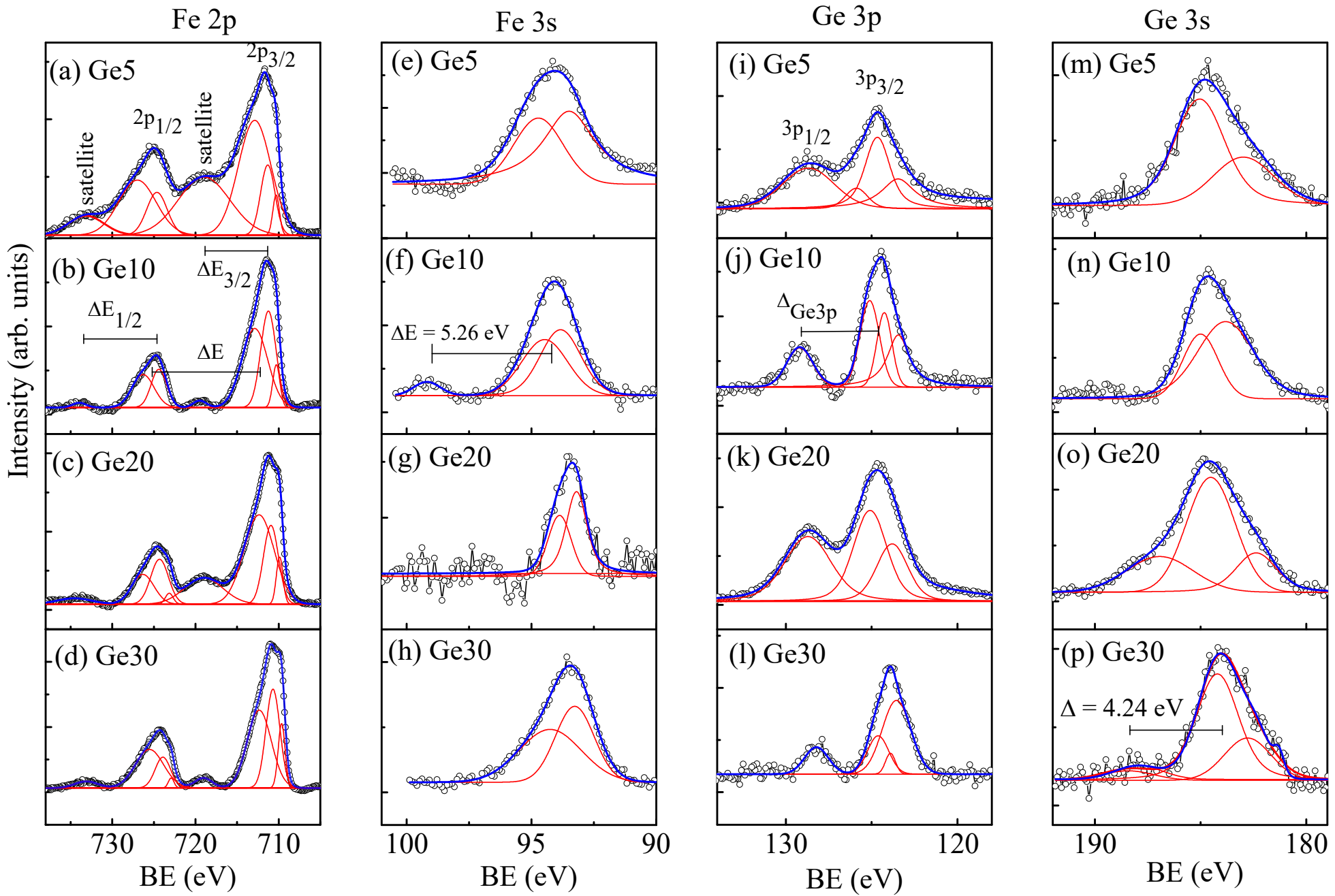

Figure 4. The deconvoluted XPS of Fe 2p (a-d), Fe 3s (e-h), Ge 3p (i-l), Ge 3s (m-p) bands for Ge doped samples.

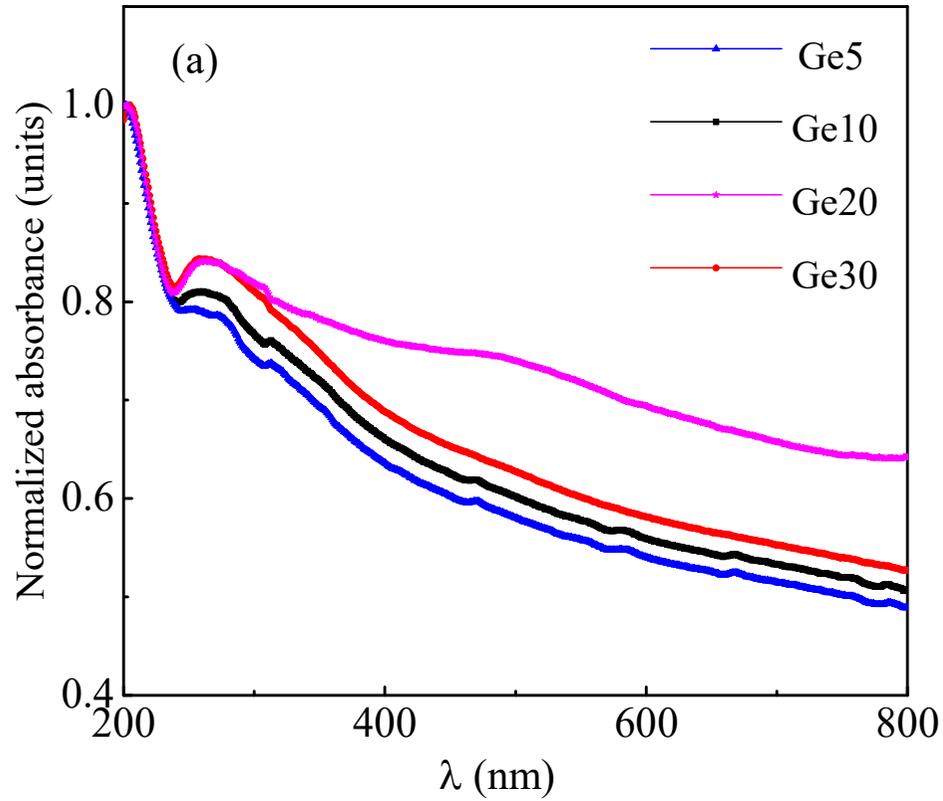
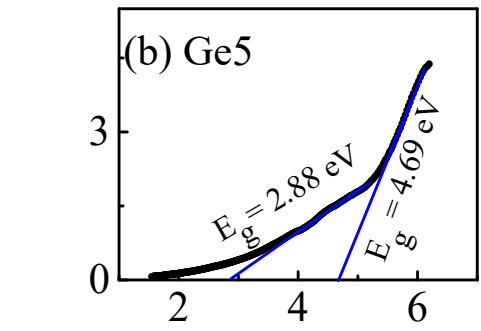
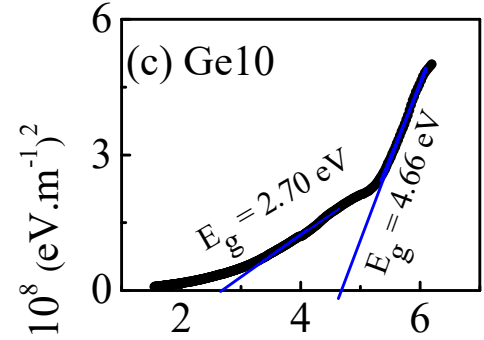
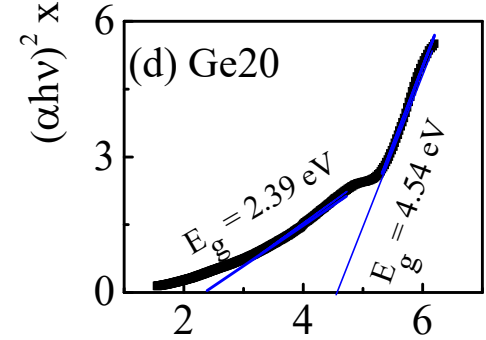
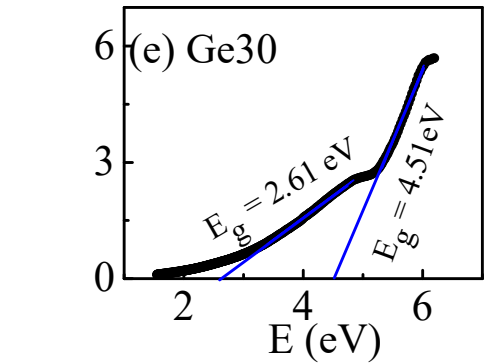
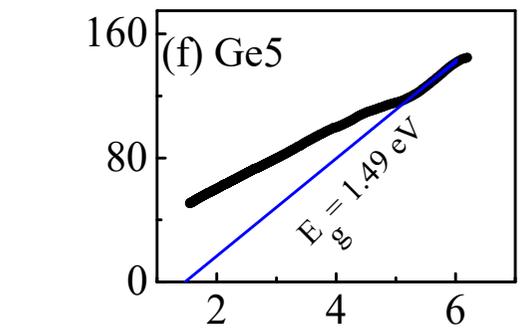
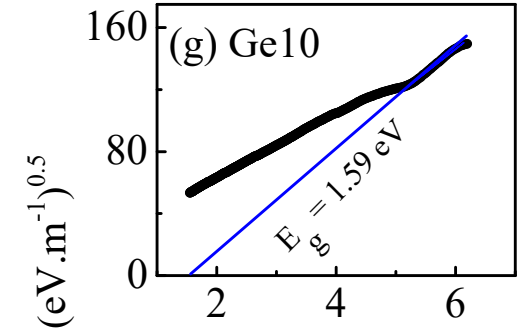
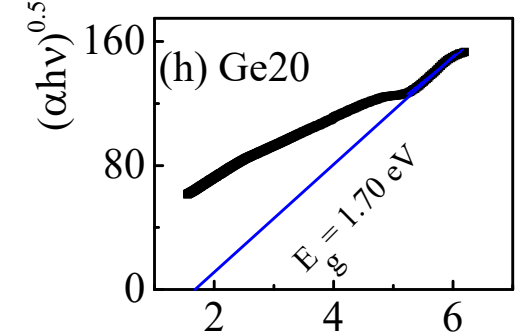
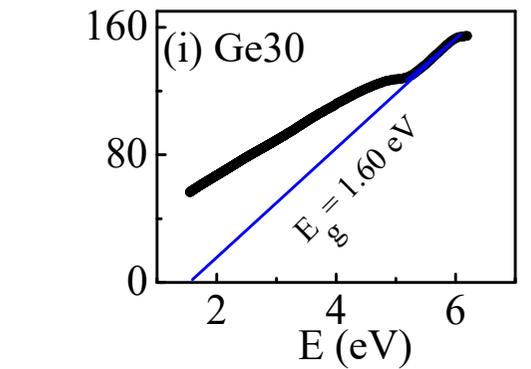

Figure 5. (a) Absorbance as a function of wavelength shown for Ge5, Ge10, Ge20 and Ge30. Tauc plots showing direct (b-e) and indirect (f-i) bang gap values for the Ge-doped hematite samples.

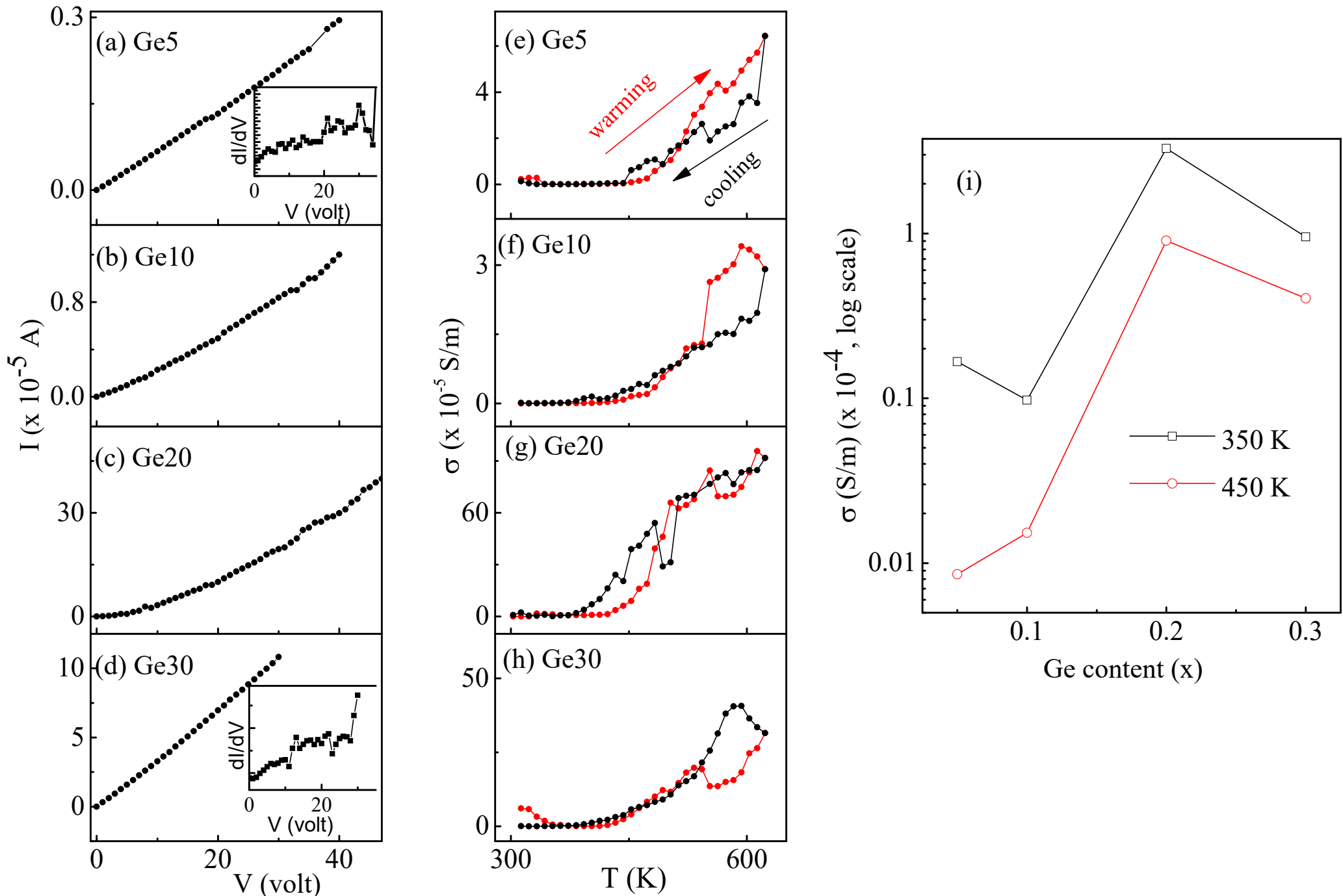

Figure 6. I-V plots of Ge doped $Fe_2O_3$ samples at 463 K (a-d) and *dI/dV* vs *V* curves for two samples (insets in (a, d)). The σ (T) curves during warming and cooling modes of measurements at 10 V (e-h). The variation of electrical conductivity (σ) with increasing dopant (Ge) concentration at 10 V (i) for measurement temperatures 350 K and 450 K.

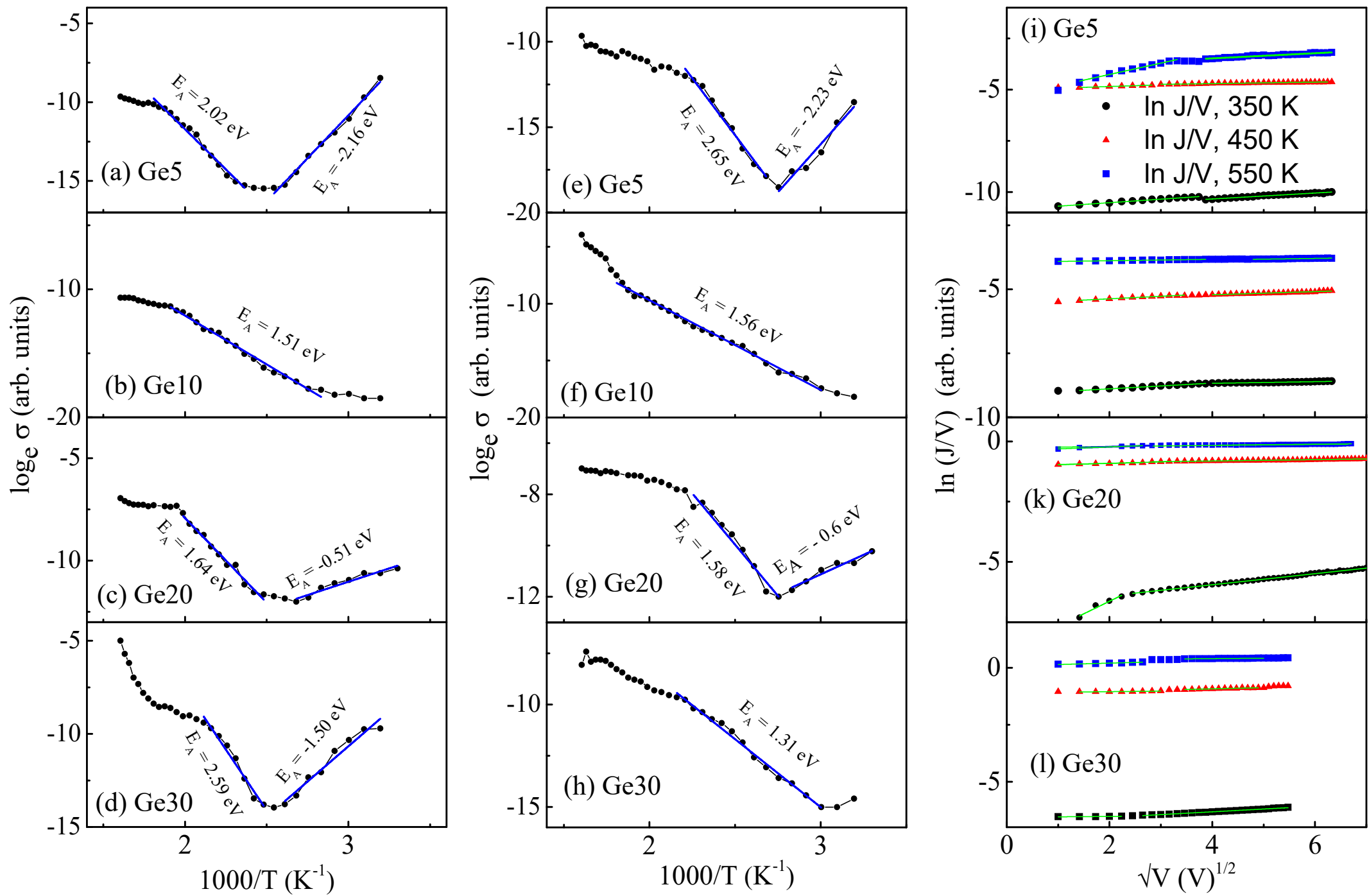

Figure 7. Fit of σ(T) data at 10 V according to Arrhenius law in warming (a-d) and cooling (e-h) modes of measurements. Analysis of I-V curves during warming mode according to Poole-Frenkel conduction mechanism at 3 temperatures (i-l).

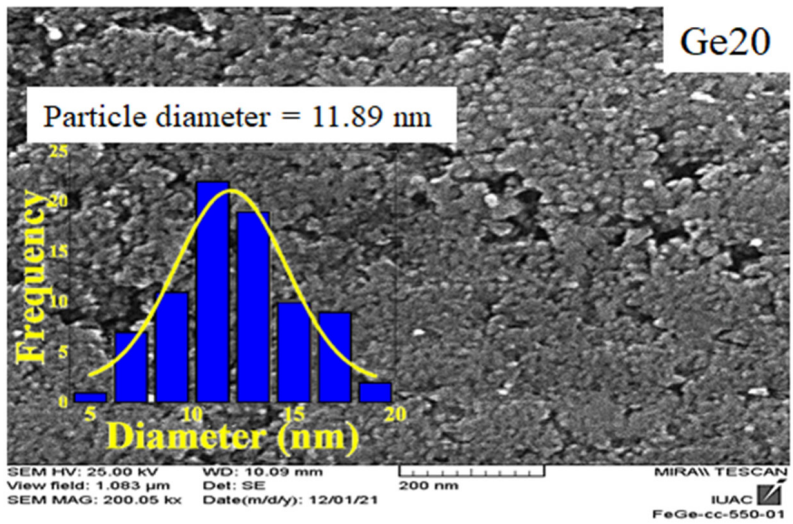

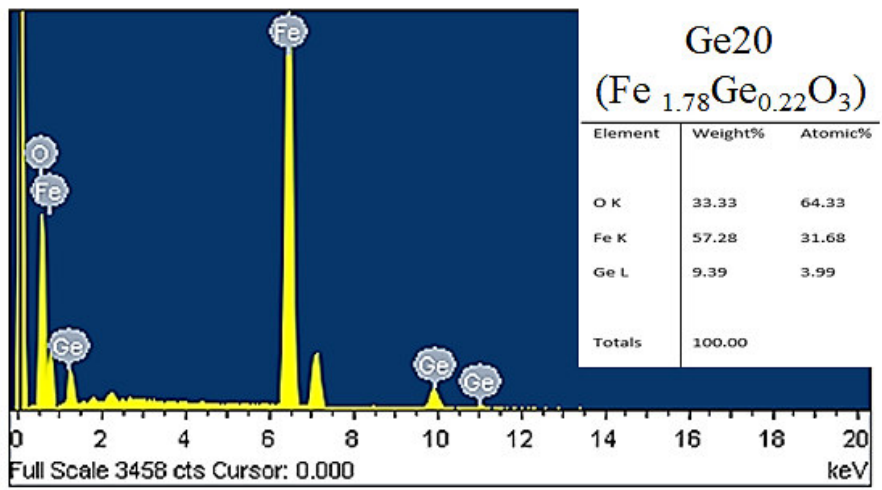

Figure S1. FESEM and EDX images of Ge20 sample showing particle distribution and sample composition.